\documentclass[a4paper,11pt]{article}
\usepackage[ams]{}
\usepackage{amssymb}
\usepackage[theorem]{}
\begin{document}
\title{Grassmann Variables in Jordan Matrix Models}
\author{Michael Rios\footnote{email: mrios4@calstatela.edu}\\\\\emph{California State University, Los Angeles}\\\emph{Mathematics Graduate Program}
\\\emph{5151 State University Drive}\\\emph{Los Angeles, CA 90032-8531}  } \date{\today}\maketitle
\begin{abstract}
Here we demonstrate the emergence of Grassmann variables in matrix models based on the exceptional Jordan algebra.  The Grassmann algebras are built naturally using the octonion algebra.  We argue the appearance of Grassmann variables solidifies the relationship between supersymmetry and triality.             
\\\\
$Keywords:$ The exceptional Jordan algebra, Grassmann variables, matrix models, supersymmetry.
\end{abstract}
\newpage
\tableofcontents
\section{Introduction}
A problem common to all matrix models based on the exceptional Jordan algebra $\mathfrak{h}_3(\mathbb{O})$ is the absence of Grassmann variables.  However, as was shown by Catto in \cite{5} it is possible to construct a Grassmann algebra using the bioctonions (split octonions) $\mathbb{C}\otimes\mathbb{O}$.  Using this result for the bioctonions we build Grassmann algebras using elements of $\mathbb{O}^2$ and $(\mathbb{C}\otimes\mathbb{O})^2$, and argue that exceptional Jordan matrix models such as the cubic matrix model \cite{1} and the $E_6$ matrix model \cite{2} do contain Grassmann variables.      
\section{Exceptional Matrix Models}
\subsection{Smolin's Cubic Matrix Model}
In \cite{1}, Smolin introduced a matrix model based on the exceptional Jordan algebra $\mathfrak{h}_3(\mathbb{O})$.  The motivation for the model stems from the $\mathfrak{h}_2(\mathbb{O})$ representation of $9 + 1$ Minkowski spacetime:
\begin{equation}
j = \left(\begin{array}{cc}a_+ & \varphi_1 \\ \overline{\varphi}_1 & a_- \\ \end{array}\right)\qquad a\in\mathbb{R}\quad\varphi\in\mathbb{O}
\end{equation}
where the $a_{\pm}$ are taken as light cone coordinates, and the eight tranverse coordinates are contained in the octonion $\varphi$.  The relevant spinors are those from the octonionic spinor representation of $\mathfrak{so}(9)$, where $\psi \in \mathbb{O}^2$.  In column form the spinors appear as:
\begin{equation}
\psi=\left(\begin{array}{cc} \overline{\varphi}_2 \\  \varphi_3 \end{array}\right)
\end{equation}
Smolin proposed a unification of the vector $j$ and spinor $\psi$ of 9 + 1 dimensional Minkowski spacetime, along with an eleventh spatial coordinate $a_3$, by embedding them in the structure of the exceptional Jordan algebra $\mathfrak{h}_3(\mathbb{O})$:
\begin{equation}
\Phi=\left(\begin{array}{cc}j & \psi \\ \psi^* & a_3 \end{array}\right)=\left(\begin{array}{ccc}a_1 & \varphi_1 & \overline{\varphi}_2 \\ \overline{\varphi}_1 & a_2 & \varphi_3 \\ \varphi_2 & \overline{\varphi}_3 & a_3 \end{array}\right)
\end{equation}
\indent The dynamics of the model are expressed in terms of matrix Chern-Simons theory rather than a matrix compactifcation of Yang-Mills theory, so that dependence on a particular background is avoided.  The degrees of freedom of the model live in $\mathcal{G}\times \mathfrak{h}_3(\mathbb{O})$, where $\mathcal{G}$ is a Lie algebra.  The action is defined using triality generators, or cycle mappings \cite{2}, which are discrete automorphisms $\rho\in F_4$ acting as:
\begin{equation}
\rho\circ\Phi = \left(\begin{array}{ccc}a_2 & \varphi_3 & \varphi_1 \\ \overline{\varphi}_3 & a_3 & \overline{\varphi}_2 \\ \overline{\varphi}_1 & \varphi_2 & a_1 \end{array}\right) 
\end{equation}
cycling the entries of the matrix, mixing up the vector, spinor and conjugate spinor of 9 + 1 Minkowski spacetime.  The action is written using the trilinear form $tr(\Phi_1, \Phi_2, \Phi_3)$ $\Phi_i\in\mathfrak{h}_3(\mathbb{O})$ as:
\begin{equation}
S = \frac{k}{4\pi}f_{IJK}tr(\Phi^I, \rho\circ\Phi^J, \rho^2\circ\Phi^K)
\end{equation}
where $f_{IJK}$ are antisymmetric structure constants of $\mathcal{G}$, and this defines Smolin's exceptional cubic matrix model.\\
\indent The complications that surface in this model involve the absence of Grassmann variables \cite{1}\cite{2} and the fact that $F_4$ is too small to recover the gauge symmetry of the standard model.  The former problem of gauge symmetry was shown to be overcome by use of the exceptional Jordan C*-algebra $\mathfrak{h}_3(\mathbb{C}\otimes\mathbb{O})$ \cite{4} which has $E_6$ symmetry \cite{2}.  In this paper, we show the problem of Grassmann variables is solved by the use of the bioctonions $\mathbb{C}\otimes\mathbb{O}$ which admit an exceptional Grassmann algebra formulation.
\subsection{$\mathbb{O}^2$ Grassmann Variables}
The construction of $\mathbb{O}^2$ Grassmann variables is based on the Grassmann representation of $\mathbb{C}\otimes\mathbb{O}$.  Elements of the bioctonionic algebra $\mathbb{C}\otimes\mathbb{O}$ can be expressed as:
\begin{equation}
\theta=\varphi_1+\textbf{i}\varphi_2\qquad \varphi\in\mathbb{O}
\end{equation}
where $\textbf{i}$ commutes with the octonions and satisfies $\textbf{i}^2=-1$.  There are two types of conjugation for the bioctonions: complex and octonionic conjugation.  For our purposes, we use octonionic conjugation which acts as:
\begin{equation}
\tilde{\theta}=\overline{\varphi}_1+\textbf{i}\overline{\varphi}_2
\end{equation}
Following Catto in \cite{5}, yet invoking octonionic conjugation for the $\tilde{\theta}$, we define the basis:
\begin{equation}
\theta_0=\frac{1}{2}(1+\textbf{i}e_7)\qquad \tilde{\theta}_0=\frac{1}{2}(1-\textbf{i}e_7)\qquad \theta_0\in\mathbb{C}\otimes\mathbb{O}
\end{equation}
\begin{equation}
\theta_j=\frac{1}{2}(e_j+\textbf{i}e_{j+3})\qquad \tilde{\theta}_j=\frac{1}{2}(-e_j-\textbf{i}e_{j+3})\qquad j=1,2,3\quad\theta_j\in\mathbb{C}\otimes\mathbb{O}
\end{equation}
where $e_k$ are octonionic units.  It is elementary to show the $\theta_j$ satisfy:
\begin{equation}
\{\theta,\theta\}=\{\theta,\tilde{\theta}\}=\{\tilde{\theta},\tilde{\theta}\}=0
\end{equation}
by recalling $\textbf{i}^2=-1$ and using the octonionic unit property $e_je_k=-e_ke_j$.  Hence, the bioctonions provide a (nonassociative) Grassmann algebra.  To show the construction is applicable to spinors $\psi\in\mathbb{O}^2$, we invoke the isomorphism:
\begin{equation}
\mathbb{O}^2\cong\mathbb{C}\otimes\mathbb{O}
\end{equation}
which at the element level amounts to:
\begin{equation}
\psi=\left(\begin{array}{cc} \overline{\varphi}_2 \\  \varphi_3 \end{array}\right)\leftrightarrow\overline{\varphi}_2+\textbf{i}\varphi_3=\theta\qquad \psi\in\mathbb{O}^2\quad\theta\in\mathbb{C}\otimes\mathbb{O}.
\end{equation}
With this correspondence in mind, we refer back to Smolin's model and see triality mappings as mixing bosonic and fermionic degrees of freedom. 
\subsection{Ohwashi's $E_6$ Matrix Model}
Ohwashi's $E_6$ matrix model is a complex extension of Smolin's cubic matrix model.  It uses matrices from $\mathfrak{h}_3(\mathbb{C}\otimes\mathbb{O})$ rather than elements of $\mathfrak{h}_3(\mathbb{O})$.  It has a compact $E_6$ symmetry and a Chern-Simons like structure.  Compact $E_6$ is defined using the cubic form as:
\begin{equation}
E_6=\{\alpha\in Iso_c(\mathfrak{h}_3(\mathbb{C}\otimes\mathbb{O}))|(\alpha \Upsilon^I,\alpha \Upsilon^J, \alpha \Upsilon^K)=(\Upsilon^I,\Upsilon^J,\Upsilon^K)\}\nonumber   
\end{equation}
where $\Upsilon^I,\Upsilon^J,\Upsilon^K\in\mathfrak{h}_3(\mathbb{C}\otimes\mathbb{O})$ satisfy $<\alpha \Upsilon^I, \alpha \Upsilon^J>=<\Upsilon^I, \Upsilon^J>$.\\  The action
\begin{equation}
S=(\rho^2\circ\Upsilon^{[I},\rho\circ\Upsilon^J, \Upsilon^{K]})f_{IJK}
\end{equation}
where [.] is weight-1 antisymmetrization on the indices, is invariant under the $E_6$ mapping:
\begin{equation}
(\alpha(\rho^2\circ\Upsilon^{[I}),\alpha(\rho\circ\Upsilon^J), \alpha\Upsilon^{K]})f_{IJK}=(\rho^2\circ\Upsilon^{[I},\rho\circ\Upsilon^J, \Upsilon^{K]})f_{IJK}=S.
\end{equation}
In \cite{2} Ohwashi divides matrices of $\mathfrak{h}_3(\mathbb{C}\otimes\mathbb{O})$ as:
\begin{equation}
\Upsilon=\left(\begin{array}{ccc}z_1 & \tilde{\xi}_1 & \xi_2 \\ \xi_1 & z_2 & \tilde{\xi}_3 \\ \tilde{\xi}_2 & \xi_3 & z_3 \end{array}\right)=\left(\begin{array}{cc}\mathcal{W} & \Psi \\ \Psi^* & v \end{array}\right)\qquad\xi\in\mathbb{C}\otimes\mathbb{O}\quad z\in\mathbb{C}
\end{equation}
where $\mathcal{W}\in\mathfrak{h}_2(\mathbb{C}\otimes\mathbb{O})$, $\Psi\in(\mathbb{C}\otimes\mathbb{O})^2$ and $v\in\mathbb{C}$.  The $v$ and $\mathcal{W}$ are taken as bosonic fields, while the $\Psi$ is intended to be a fermionic field.\\
\subsection{$(\mathbb{C}\otimes\mathbb{O})^2$ Grassmann Variables}
\indent A closer analysis of the $\Psi\in(\mathbb{C}\otimes\mathbb{O})^2$, in light of the Grassman algebra of section 2.2, shows the $\Psi$ to indeed transform as fermionic fields.  The $\Psi$ are written in column form as:
\begin{equation}
\Psi=\left(\begin{array}{cc} \xi_2 \\ \xi_3 \end{array}\right).
\end{equation}
In complex form, they are written as:
\begin{equation}
\Psi=\xi_2+\textbf{i}\xi_3
\end{equation}
where the $\textbf{i}$ commutes with the $\xi$'s and satisfies $\textbf{i}^2=-1$.  Conjugation is:
\begin{equation}
\Psi^*=\tilde{\xi}_2+\textbf{i}\tilde{\xi}_3
\end{equation}
Using the $\theta_j$'s from the basis of $\mathbb{C}\otimes\mathbb{O}$ defined in section 2.2, we re-express the $\Psi$'s as:
\begin{equation}
\Psi=\theta_{j_1}+\textbf{i}\theta_{j_2}
\end{equation}
Using the Grassmann properties of the $\theta_j$, such as:
\begin{equation}
\theta_j^2=0\quad\theta_j\theta_k=-\theta_k\theta_j
\end{equation}
the $\Psi$ are seen to satisfy the Grassmann algebra properties:
\begin{equation}
\{\Psi,\Psi\}=\{\Psi,\Psi^*\}=\{\Psi^*,\Psi^*\}=0.
\end{equation}
Thus the complex representation of $\Psi\in(\mathbb{C}\otimes\mathbb{O})^2$, with a suitable basis, produces a Grassmann algebra.  The $\Psi$ can be regarded as fermionic fields after all, so Ohwashi's view of elements of $\mathfrak{h}_3(\mathbb{C}\otimes\mathbb{O})$ as matrices of bosonic and fermionic fields is supported.  As with the cubic matrix model of Smolin, triality generators (cycle mappings) of $\mathfrak{h}_3(\mathbb{C}\otimes\mathbb{O})$ mix these fermionic and bosonic fields in the $E_6$ model. 
\section{Conclusion}
In this paper we have shown that it is possible to have Grassmann variables in matrix models based on the exceptional Jordan algebra.  This makes the cubic matrix model and the $E_6$ matrix model serious candidates for nonperturbative descriptions of M-theory.  Octonionic Grassmann algebras signal a deeper algebraic structure hidden within theories of quantum gravity, as P. Ramond has explained \cite{6}.  This deeper structure surely includes the normed division algebras $\mathbb{R},\mathbb{C},\mathbb{H}$ and $\mathbb{O}$, as well as the exceptional Lie algebras and projective spaces built \cite{3} from them.\\
\indent The geometric framework arising from the algebraic structures seems to take us beyond the realm of manifolds and even noncommutative geometry.  This is because structures built from $\mathbb{O}$ exhibit nonassocitivity.  In result, we have hybrid C*-algebras such as the exceptional Jordan C*-algebra $\mathfrak{h}_3(\mathbb{C}\otimes\mathbb{O})$ \cite{4} that are commutative but nonassociative.  If we imagine examining the spectrum of such a C*-algebra and attempt to interpret the commutative geometry via the theorem of Gel'fand, we fall short because of nonassociativity.  Thus, we have geometries that are not exactly commutative geometries and noncommutative geometries, and conventional mathematical techniques are insufficient.  We must extend the spectral tools of noncommutative geometry to account for nonassociative maximal ideals spaces.  Fortunately, there are alternative routes to exploring the geometries of algebras such as $\mathfrak{h}_3(\mathbb{C}\otimes\mathbb{O})$, as will be shown in the author's next paper \cite{7}.  Where conventional mathematics wanes, D-brane technology excels.\\\\
Special thanks to: Lee Smolin and Yuhi Ohwashi for their helpful discussions on the need for Grassmann variables in exceptional Jordan matrix models.


\begin{thebibliography}{-label}
\bibitem[1]{1}L. Smolin, \textit{The Exceptional Jordan Algebra and the Matrix String}, \texttt{hep-th/0104050}.
\bibitem[2]{2}Y. Ohwashi, \textit{$E_6$ Matrix Model}, \texttt{hep-th/0110106}.
\bibitem[3]{3}J. C. Baez, \textit{The Octonions}, \texttt{math.RA/0105155}.
\bibitem[4]{4}J. D. M. Wright, \textit{Jordan C*-algebras}, Mich. Math. J. 24 (1977), 291-302.
\bibitem[5]{5}S. Catto, \textit{Exceptional Projective Geometries and Internal Symmetries}, \texttt{hep-th/0302079}.
\bibitem[6]{6}P. Ramond, \textit{Algebraic Dreams}, \texttt{hep-th/0112261}.
\bibitem[7]{7}M. Rios, \textit{The Geometry of Jordan Matrix Models}, (unpublished).
\end{thebibliography}
\end{document}